\documentclass[osajnl,twocolumn,showpacs,superscriptaddress,10pt]{revtex4-1} 
\usepackage{amsmath,amssymb,graphicx}
\newcommand{\mic}{~\mu {\rm m}}

\begin{document}

\title{Experimental observation of long-wavelength dispersive wave generation induced by self-defocusing nonlinearity in BBO crystal}

\author{Binbin Zhou}
\author{Morten Bache}
\email{moba@fotonik.dtu.dk}
\affiliation{DTU Fotonik, 
Technical University of Denmark,
  Bld. 343, DK-2800 Kgs. Lyngby, Denmark.\\}

\begin{abstract}We experimentally observe long-wavelength dispersive waves generation in a BBO crystal. A soliton was formed in normal GVD regime of the crystal by a self-defocusing and negative nonlinearity through phase-mismatched quatradic interaction. Strong temporal pulse compression confirmed the formation of soliton during the pulse propagation inside the crystal. Significant dispersive wave radiation was measured in the anomalous GVD regime of the BBO crystal. With the pump wavelengths from 1.24 to 1.4 $\mu$m, tunable dispersive waves are generated around 1.9 to 2.2 $\mu$m. The observed dispersive wave generation is well understood by simulations.

\end{abstract}


\maketitle 

Fiber-based dispersive wave (DW) induced by solitons has been intensively investigated. It plays an essential role in fiber supercontiuum generation  \cite{2003Skryabin,2006Dudley}. New spectral components on either the short-wavelength side or the long-wavelength side of the pump wavelength are possible by this technique with proper fiber dispersion control \cite{2003Tartara,2010Chang,2013Mak,2008Falk,2010Kolesik,2013Yuan}. The normally small core size of the fiber, however, limits the propagating pulse energy; especially when generating DW in the long-wavelength side of the pump and multiple zero-dispersion wavelengths (ZDWs) are needed. The direct use of bulk nonlinear materials is promising for high pump energy. Most bulk nonlinear materials have normal group velocity dispersion (GVD) in the near-infrared (NIR) pump wavelength; therefore, when a conventional Kerr (self-focusing and positive) nonlinearity is employed, the study of DW generation easily falls into another dilemma: filamentation likely kicks in. Which complicates the whole process, and greatly impedes its application \cite{1990Golub,1996Nibbering,2005Kolesik}.

A negative (self-defocusing) nonlinearity, e.g. the one available from a phase-mismatched quadratic process, could open a new window for DW research for the mostly used NIR pump wavelengths. The flipped sign of nonlinearity means that instead of filamentations, the generation of soliton is possible in the normal GVD regime of the bulk material in the NIR; which will then be able to stimulate efficient DW emission in the anomalous GVD regime (longer wavelength) of the crystal. Furthermore, much higher pulse energy could be supported due to the absence of filamentation. The excitation and utilization of NIR soliton in the normal GVD regime by the phase-mismatched quadratic process have been reported in various nonlinear crystals \cite{1999Liu,2002Ashihara,2006Moses,2007Moses,2012Zhou}; especially in the most frequently used BBO crystal, by which soliton compression down to few optical cycle has been realized \cite{2006Moses}. However, despite the success of exciting self-defocusing temporal soliton in the BBO crystal, and the numerical prediction of the existence of such a DW emission \cite{2010Bache}, the formation of DW emission in the anomalous GVD regime of this crystal has so far never been observed experimentally. It was only recently observed directly for the first time in cascaded nonlinear media \cite{2015Zhou}: The strong few-cycle near-IR self-defocusing soliton self-compression observed in a bulk lithium niobate crystal in \cite{2012Zhou} was confirmed to give rise to a resonant transfer of energy to a femtosecond mid-IR dispersive wave \cite{2015Zhou}.

\begin{figure}[tb]
  \centering{
  \includegraphics[width=\linewidth]{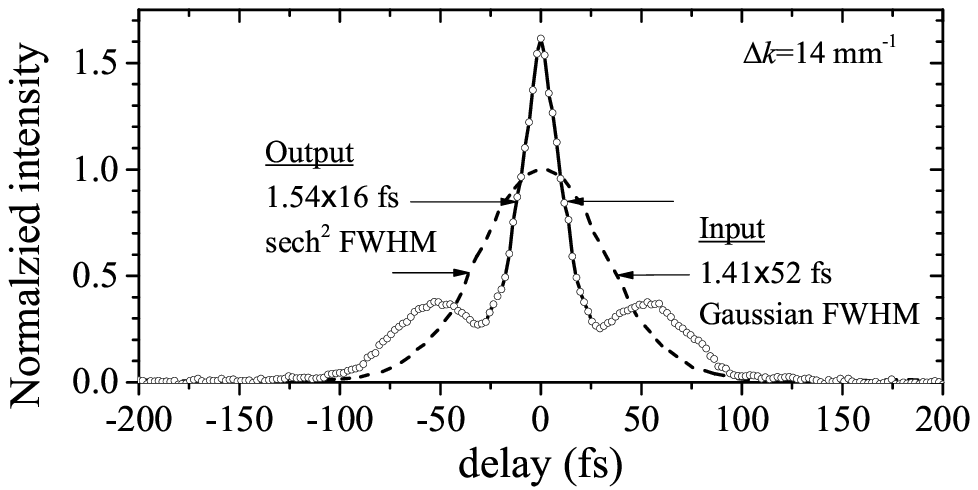}
  \includegraphics[width=\linewidth]{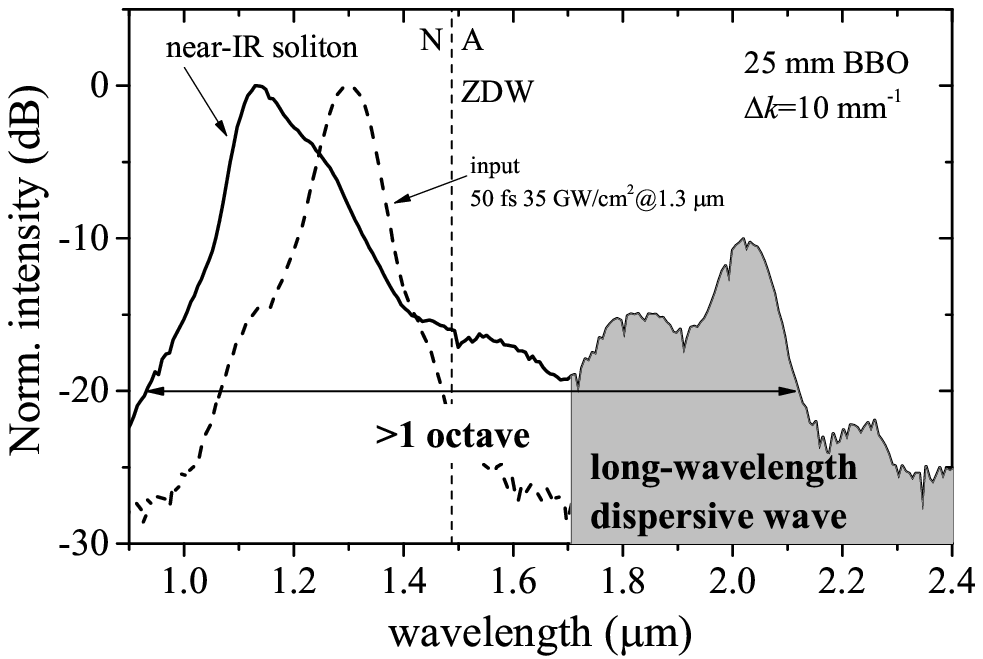}
  \vspace{0mm}
  } \caption{Experimentally observed soliton self-compression and dispersive wave formation in a 25 mm BBO pumped at $\lambda_1=1.3\mic$ with peak intensity was $35~{\rm GW/cm}^2$. Top: Intensity AC traces for the input pulse (corresponding to a 52 fs FWHM Gaussian pulse) and the output pulse, showing a 16 fs self-compressed self-defocusing soliton. The output was measured for $\Delta k=14~\rm mm^{-1}$, but similar traces were found for other values of $\Delta k$. Bottom: The input and output spectra for $\Delta k=10 ~\rm mm^{-1}$: An over 1 octave supercontinuum has formed, and the near-IR soliton, located in the normal ('N') dispersion regime resonantly couples to a long-wavelength dispersive wave located in the anomalous ('A') dispersion regime.
  }\label{fig:example}
\end{figure}

In this letter, we will demonstrate the first experimental observation results of DW generation by such a negative nonlinearity in BBO crystal. DW emission from 1.9 to 2.2 $\mu$m are measured in a 25-mm-long BBO crysal with single pump wavelengths from 1.24 to 1.40 $\mu$m.

For the experiment, a piece of 25-mm-long BBO crystal with a $10\times7~{\rm mm^2}$ aperture (cut with $\theta=21^{\circ}$, $ \phi=-90^\circ $) is used for the phase-mismatched quadratic nonlinear process. The wavelength tunable output laser from a 1 kHz commercial OPA system are pumped through the crystal (beam spot size 1.8 mm FWHM). The pulse duration of the laser pulses from the OPA was around 50 fs (see below) and were kept slightly negatively chirped. The phase-mismatch was tuned by rotating the external crystal angle, and the pump intensity was adjusted by neutral density filters. An InGaAs CCD-based spectrometer is employed to monitor the spectrum from 870 up to 2500 nm. The pulse duration of the output pulses were checked by a intensity autocorrelator with a 100 $\mu$m thick BBO SHG crystal, and were typically $\simeq 50$ fs FWHM and Gaussian shaped. The bandwidth was measured to be around $60$ nm, so the pulses are slightly above transform limited duration (a linear chirp parameter $|C|=0.9$ is deduced, which is beneficial for soliton formation as long as it is negative \cite{2012Zhou}). The basic experimental setup is similar to what was used in \cite{2006Moses}; and similarly, we were able to observe clear temporal pulse compression under various pump intensities and phase-mismatch values. Because there is no dispersion compensation involved, the pulse self-compression is a clear indication that high order soliton was excited due to the interplay of the positive material GVD and a negative quadratic nonlinearity.

The top plot in Fig. \ref{fig:example} shows the measured autocorrelation traces for the direct input and the output pulses ($\Delta k$ = 14$\rm mm^{-1}$ and $35~{\rm GW/cm}^2$). Despite the existence of uncompressed pedestal structure, which is common to see with pulse compression by soliton, the main part of the pulse was compressed down by nearly 3 folds: from 52 fs FWHM (using the Gaussian de-convolution factor) down to about 16 fs FWHM (using the sech$^2$ de-convolution factor). The output spectrum shown in the lower plot reveals that the formation of temporal soliton around the NIR pump wavelength indeed is accompanied by DW formation in the long-wavelength anomalous GVD regime of the BBO crystal. The peak is found only 10 dB below the soliton maximum, confirming that the energy transfer is quite substantial. The bandwidth of the supercontinuum is over 1 octave. 


\begin{figure}[tb]
  \centering{
    \includegraphics[width=\linewidth]{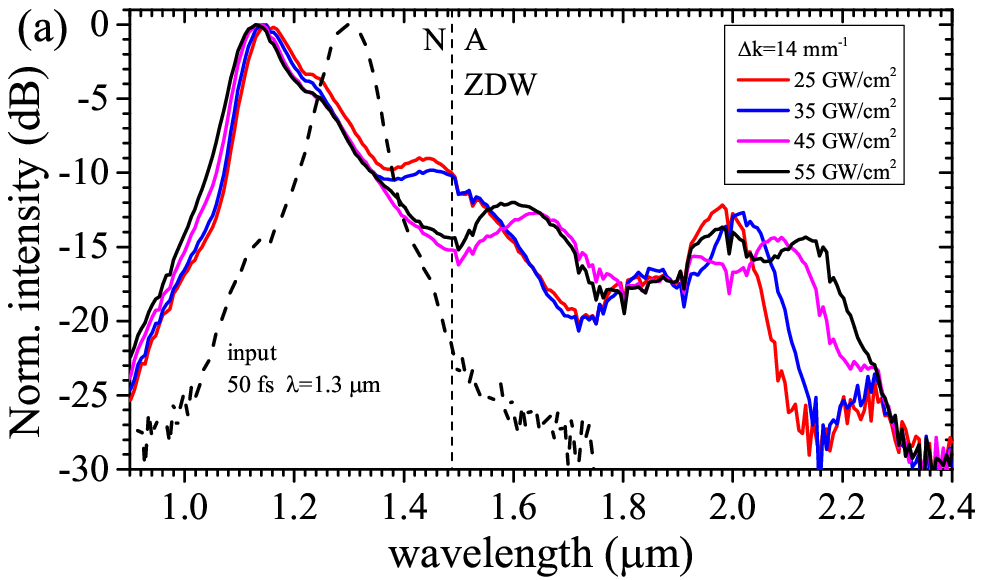}
    \includegraphics[width=\linewidth]{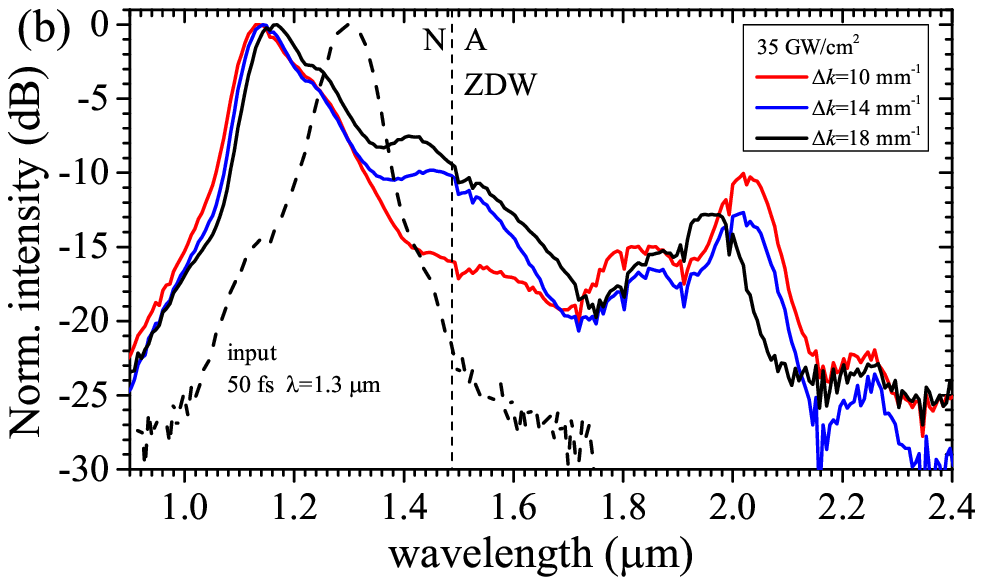}
  \vspace{0mm}
  } \caption{Experimentally recorded spectra for $\lambda_1=1300$ nm. (a) Crystal tuning angle is fixed ($\Delta k$ = 14$\rm mm^{-1}$) and the pump intensities increase from 25 to $55~{\rm GW/cm}^2$. The BBO ZDW is indicated by a vertical line. (b) Pump intensity is fixed ($35~{\rm GW/cm}^2$) and $\Delta k$ is tuned.
}\label{fig:1300}
\end{figure}

We were able to observe the generation of DW way beyond the zero dispersion wavelength of the crystal under various conditions. Fig. \ref{fig:1300}(a) shows the input pump spectrum at 1300 nm and the measured output spectra when the crystal is angle-tuned for $\Delta k$ = 14$\rm mm^{-1}$ and pumped by increased intensity. The spectral broadening is tremendous (more than one octave at -20dB level), and the significant secondary spectral peaks near 2.0 to 2.1 $\mu$m are clear sign of DW generation in the linear regime. With increased pump intensities, the formed DW slightly shifts to the red. We can also notice that the central wavelength of the NIR pump experiences significant amount of blue shift, and  NIR peaks move a bit further to the blue side with higher pump intensity. The large blue shift of the original pump spectrum actually holds the DW generation from going much further into the longer wavelength by rectifying the phase-mismatch $\Delta k$; the $\Delta k$ value will be significantly increased with reduced pump wavelength under the same pump angle inside the BBO crystal, and a large $\Delta k$ could pull the DW to shorter wavelength. The results agree well with the numerical predictions. Fig. \ref{fig:1300}(b) shows the dynamic of the DW generation under shifting phase mismatch values. The DW emission is gradually strengthened with the phase-mismatch value decreases from 18 to 14 $\rm mm^{-1}$.

\begin{figure}[tb]
  \centering{
  \includegraphics[height=3.1cm]{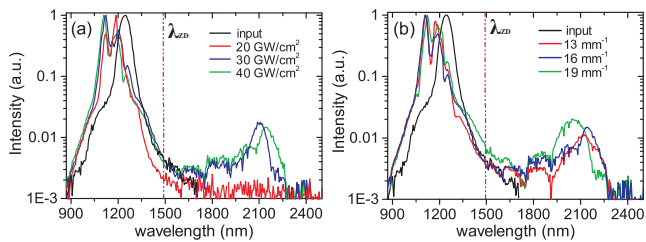}
  \vspace{0mm}
  } \caption{(a) The input spectra at 1240 nm and the output spectra when the crystal tuning angle is fixed ($\Delta k$ = 16$\rm mm^{-1}$) and the pump intensities increase from 20 to $40~{\rm GW/cm}^2$. (b) Output spectra when the pump intensity is fixed at $40~{\rm GW/cm}^2$ and $\Delta k$ is tuned.
}\label{fig:1240}
\end{figure}

\begin{figure}[tb]
  \centering{
  \includegraphics[height=3.1cm]{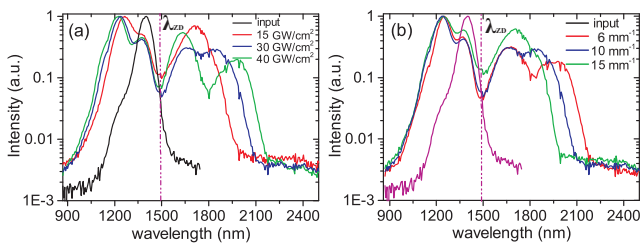}
  \vspace{0mm}
  } \caption{(a) The input spectra at 1400 nm and the output spectra when the crystal tuning angle is fixed ($\Delta k$ = 10$\rm mm^{-1}$) and the pump intensities increase from 15 to $40~{\rm GW/cm}^2$. (b) Output spectra when the pump intensity is fixed at $30~{\rm GW/cm}^2$ and $\Delta k$ is tuned.
}\label{fig:1400}
\end{figure}

DW generation at other pump wavelengths were also observed. Fig. \ref{fig:1240} and Fig. \ref{fig:1400} show relevant experimental results at the pump wavelength of 1240 nm and 1400 nm, respectively. In general, when the pump NIR wavelength is closer to the ZDW, the generation of DW is more efficient, which can be well understood: the efficiency of DW generation depends on the overlapping between the soliton spectrum tail and the DW phase matching point; longer pump wavelength means more energy coupling from the soliton into the DW.

In conclusion, we report the experimental observation of long-wavelength DW generation by negative nonlinearity in BBO crystal. With the formation of self-defocusing soliton in the NIR pump wavelength of 1.24 to 1.40 $\mu$m, DW is observed around 1.9 to 2.2 $\mu$m, in the anomalous GVD regime of the crystal. Significant blue shift of the NIR pump spectrum is also observed, which at some extend prohibit the DW radiation in further longer wavelengths. The generated DWs are tunable by the angle-tuning of phase mismatch or pump intensity. Such technique can be easily applied to other bulk quadratic crystals and enable DW generation in different wavelength ranges \cite{2011Bache,2015Zhou}.


\begin{thebibliography}{99}


\bibitem{2003Skryabin} D. V. Skryabin, F. Luan, J. C. Knight, and P. St. J. Russell, Science {\bf 301,} 1705-1708 (2003).
\bibitem{2006Dudley} J. M. Dudley, G. Genty, and S. Coen, Rev. Mod. Phys. {\bf 78,} 1135-1184 (2006).
\bibitem{2003Tartara} L. Tartara, I. Cristiani, and V. Degiorgio, Appl. Phys. B {\bf 77,} 307-311 (2003).
\bibitem{2010Chang} G. Chang, L. J. Chen, and F. X. K{\"a}rtner, Opt. Lett. {\bf 35,} 2361-2363 (2010).
\bibitem{2013Mak} K. F. Mak, J. C. Travers, P. H{\"o}lzer, N. Y. Joly, and P. St. J. Russell, Opt. Express {\bf 21,} 10942-10953 (2013).
\bibitem{2008Falk} P. Falk, M. H. Frosz, O. Bang, L. Thrane, P. E. Andersen, A. O. Bjarklev, K. P. Hansen, and J. Broeng, Opt. Lett. {\bf 33,} 621-623 (2008).
\bibitem{2010Kolesik} M. Kolesik, L. Tartara, and J. V. Moloney, Phys. Rev. A {\bf 82,} 045802 (2010).
\bibitem{2013Yuan} J. H. Yuan, X. Z. Sang, Q. Wu, C. X. Yu, K. R. Wang, B. B. Yan, X. W. Shen, Y. Han, G. Y. Zhou, Y. Semenova, G. Farrell, and L. T. Hou, Laser Phys. Lett. {\bf 10,} 045405 (2013).
\bibitem{1990Golub} I. Golub, Opt. Lett. {\bf 15,} 305-307 (1990).
\bibitem{1996Nibbering} E. T. J. Nibbering, P. F. Curley, G. Grillon, B. S. Prade, M. A. Franco, F. Salin, and A. Mysyrowicz, Opt. Lett. {\bf 21,} 62-64 (1996).
\bibitem{2005Kolesik} M. Kolesik, E. M. Wright, and J. V. Moloney, Opt. Express {\bf 13,} 10729-10741 (2005).
\bibitem{1999Liu} X. Liu, L. Qian, and F. W. Wise, Opt. Lett. {\bf 24,} 1777-1779 (1999).
\bibitem{2002Ashihara} S. Ashihara, J. Nishina, T. Shimura, and K. Kuroda, J. Opt. Soc. Am. B {\bf 19,} 2505-2510 (2002).
\bibitem{2006Moses} J. Moses and F. W. Wise, Opt. Lett. {\bf 31,} 1881-1883 (2006).
\bibitem{2007Moses} J. Moses, E. Alhammali, J. M. Eichenholz, and F. W. Wise, Opt. Lett. {\bf 32,} 2469-2471 (2007).
\bibitem{2012Zhou} B.B. Zhou, A. Chong, F.W. Wise and M. Bache, Phys. Rev. Lett. {\bf 109,} 043902 (2012).
\bibitem{2010Bache} M. Bache, O. Bang, B.B. Zhou, J. Moses, and F. W. Wise, Phys. Rev. A {\bf 82,} 063806 (2010).
\bibitem{2015Zhou} B.B. Zhou, H.R. Guo, and M. Bache, Opt. Express {\bf 23,} 6924-6936 (2015).
\bibitem{2011Bache} M. Bache, O. Bang, B.B. Zhou, J. Moses, and F. W. Wise, Opt. Express {\bf 19,} 22557-22562 (2011).

\end{thebibliography}
\end{document}